\documentclass{emulateapj}
\slugcomment{to be published by the Astrophysical Journal}

\usepackage{natbib}

\usepackage{graphicx}
\usepackage{rotating}

\shorttitle{DM Halos and Bars: Varying Gas Fraction and Resolution}
\shortauthors{Villa-Vargas, Shlosman and Heller}

\begin{document}

\newcommand{\ps}{$\Omega_\mathrm{b}$}  
\newcommand{\Atwo}{$A_2$}  
\newcommand{\Atworb}{$A_{2\mathrm{b}}$}  
\newcommand{\Aone}{$A_{\rm 1,z}$} 
\newcommand{\Jd}{$J_{\rm d}$}  
\newcommand{\Jh}{$J_{\rm h}$}  
\newcommand{\dJ}{$\dot{J}$} 
\newcommand{\dJd}{$\dot{J}_{\rm d}$} 
\newcommand{\dJh}{$\dot{J}_{\rm h}$} 
\newcommand{\Jdin}{$J_{\mathrm{d,in}}$}  
\newcommand{\Jdout}{$J_{\mathrm{d,out}}$}  
\newcommand{\Jhin}{$J_{\mathrm{h,in}}$}  
\newcommand{\Jhout}{$J_{\mathrm{h,out}}$}  
\newcommand{\Rcr}{$R_\mathrm{cr}$}  
\newcommand{\Rbar}{$R_\mathrm{b}$}  
\newcommand{\Rd}{$R_\mathrm{d}$}  
\newcommand{\fg}{$f_\mathrm{g}$}  
\newcommand{\egrav}{$\epsilon_\mathrm{grav}$}  
\newcommand{\mcmc}{$M_\mathrm{\rm CMC}$}  

\def\gtorder{\mathrel{\raise.3ex\hbox{$>$}\mkern-14mu
    \lower0.6ex\hbox{$\sim$}}}
\def\ltorder{\mathrel{\raise.3ex\hbox{$<$}\mkern-14mu
    \lower0.6ex\hbox{$\sim$}}}

\title{DARK MATTER HALOS AND EVOLUTION OF BARS IN DISK GALAXIES:
VARYING GAS FRACTION AND\\ GAS SPATIAL RESOLUTION}

\author{
Jorge Villa-Vargas\altaffilmark{1},
Isaac Shlosman\altaffilmark{1}
and
Clayton Heller\altaffilmark{2}
}
\altaffiltext{1}{
Department of Physics and Astronomy,
University of Kentucky,
Lexington, KY 40506-0055,
USA
}
\altaffiltext{2}{
Department of Physics,
Georgia Southern University,
Statesboro, GA 30460,
USA
}

\begin{abstract}
We conduct numerical experiments by evolving gaseous/stellar disks embedded in
live dark matter halos aiming at quantifying the effect of gas spatial resolution and
gas content on the bar evolution. Three model sequences have been constructed using
different resolution, and gas fraction has been varied along each sequence within 
the range of \fg$= 0\%-50\%$, but keeping the disk and halo properties unchanged. 
We find that the spatial resolution becomes important 
with an increase in the gas content.  For the higher resolution model sequences,
we observe a bimodal behavior in the bar evolution with respect to the gas fraction, 
especially during the secular phase of this evolution. The switch from the gas-poor
to gas-rich behavior is abrupt and depends on the resolution used, being
reasonably confined to \fg$\sim 5\%-12\%$.
The diverging evolution has been observed in nearly all basic parameters characterizing
bars, such as the bar strength, central mass concentration, bar vertical
buckling amplitude, bar size, etc. We find that the presence of the gas component
severely limits the bar growth and affects its pattern speed evolution. Gas-poor
models display rapidly decelerating bars, while gas-rich models exhibit bars
with constant or even slowly accelerating tumbling. We also find that the gas-rich
models have bar corotation (CR) radii within the disk at all times, in contrast with
gas-poor and purely stellar disks. In addition, the CR-to-bar size ratio is less
than 2 for gas rich-models. Next, we have confirmed that the disk angular momentum
within the CR remains unchanged in the gas-poor models, as long as the CR stays
within the disk, but experiences a sharp drop before leveling off in the gas-rich 
models. Finally, we discuss a number of observed correlations between various
parameters of simulated bars, such as between the bar sizes and the gas fractions,
between the bar strength and the buckling amplitude, the bar strength and its size,
etc. 
\end{abstract}


\section{Introduction}

Many issues related to the formation and evolution of galactic bars remain
unsettled (e.g., Villa-Vargas, Shlosman \& Heller 2009, hereafter Paper~I).
In Paper~I, we have revisited the properties of barred disks
embedded in the dark matter (DM) halos, with an emphasis on the angular
momentum redistribution in {\it collisionless} systems. Here we attempt
to understand some effects of the gas component on the bar evolution.
In a follow-up work, we investigate the effect of varying the halo parameters
in the gaseous/stellar disk models.

Due to its viscous and dissipative nature, the gas can influence 
the stellar component in the disk, well beyond
its observed mass fraction (e.g., Shlosman \& Noguchi 1993; Friedli \& Benz 1993;
Heller \& Shlosman 1994; Berentzen et al. 1998; Bournaud \& Combes 2002;
Berentzen et al. 2007; Curir et al. 2007). The gas dissipation, especially 
in the bar's presence, triggers the 
mass redistribution in the disk. A substantial fraction of barred orbits 
are self-intersecting, which while irrelevant for stars has a strong effect on
the gas. The gas populating such orbits is short-lived, resulting in  
shocks and loss of rotational support. Different viscosities in gas and 
stellar `fluid' lead to a delayed response to the gravitational torques and 
to an exchange of angular momentum between these components. The loss of angular 
momentum in the gas induces a flow down to the inner disk, forming 
a central mass concentration (CMC), and possibly forming and fueling the central 
supermassive black hole (SBH) via nested bars mechanism (e.g., Shlosman et 
al. 1989, 1990; Pfenniger \& Norman 1990; Friedli \& Martinet 1993; Knapen et al.
1995; Begelman \& Shlosman 2009; Hopkins \& Quataert 2010), and 
additional processes closer to the SBH (e.g., Shlosman 1999; Hopkins et al. 2010). 
The gas is capable of 
modifying the disk orbital structure (e.g., Berentzen et al. 1998). 
Moreover, the cold gas is inherently clumpy 
which effectively heats up the stellar `fluid' and populates the disk with 
an increased fraction of chaotic orbits, resulting in a decay of the bar 
strength (e.g., Shlosman \& Noguchi 1993; Berentzen et al. 2007). 
The vertical buckling instability in the bar is progressively damped by
the two-fluid gas-star interaction (e.g., Berentzen et al. 2007).

Formation and evolution of stellar bars is intricately related to the 
redistribution of angular momentum, $J$, in the disk-halo system. This means the
presence of sources and sinks of angular momentum. Sellwood (1981) has shown that
the former reside in the disk and the latter in the DM halo (see also
Debattista \& Sellwood 1998; Athanassoula 2002). Interactions of sources
and sinks of $J$ are dominated by resonances (Lynden-Bell \& Kalnajs 1972; 
Athanassoula 2002; Martinez-Valpuesta, Shlosman \& Heller 2006; Paper~I). Details 
of these resonance interactions are not fully understood and quantified. This is
especially true in the presence of the gas component, as the above cited works
are based on a purely collisionless modeling.

In this paper we analyze the effect of the gaseous component on the 
evolution of galactic bars and take a numerical approach. In particular, we
vary the gas content in the disk and advance models with various
numerical resolution, but keep the stellar disk and DM halo paramaters 
unchanged, using the Standard Model of Paper~I. We are especially interested 
in how the gas presence and the 
associated numerical resolution affect the basic parameters of a stellar
bar, as well as the angular momentum transfer process in a 
barred disk and in a disk-halo system. For this purpose, we perform a 
detailed comparison of models with gas to the standard collisionless model 
published in Paper~I.

 
\section{Numerics and Modeling}
\label{numerics}

We use the hybrid $N$-body and Smooth Particle Hydrodynamics (SPH) FTM-4.4 
code (e.g., Heller \& Shlosman 1994; Heller, Shlosman \& Athanassoula 2007;
Romano-Diaz et al. 2009) to evolve the stellar and gaseous disks embedded in the
DM halos. The gravitational forces are calculated using the FalcON routine 
(Dehnen 2002) which scales as $O(N)$. The adopted units are the same as 
in Paper~I: the units of mass and distance are taken as $10^{11}~{\rm M_\odot}$
and 10~kpc respectively. This makes the unit of time equal 
to $4.7\times 10^7$~yr, when $G = 1$, and the velocity unit
$208~~{\rm km~s^{-1}}$. The gravitational softening is $\epsilon_{\rm grav} = 
0.016$ for stars and DM particles. For the gas particles we use a dynamical 
softening. The gravitational softening is set to the smoothing length 
unless the smoothing length falls below the fixed limiting value
$\epsilon_{\rm dyn}$.
The models consist of a stellar disk with  $N_{\rm *} = 2\times 10^5$, 
a gas disk with $N_{\rm gas} = 4\times 10^4$ and DM halo with 
$N_{\rm DM} = 10^6$ collisionless particles. 
Models were evolved for about a Hubble time, $\Delta t=270$ in 
the adopted units, which translates to 12.7~Gyr.

During the model evolution, we have routinely observed the formation of very 
compact accumulations of gas particles in the central disk region. In the models
with a substantial gas component we have observed the formation of secondary
gaseous bars which decoupled from the large-scale bars and contracted 
subsequently to spatial scales where insufficient
resolution resulted in flattened `blobs' a few softening lengths in radius,
in agreement with simulations of Englmaier \& Shlosman (2004).
Besides the gradual capture of additional gas particles, the morphological 
structure of the blob has evolved very little and was beyond the resolution of our 
models. For the sake of shortening the computational time, we have replaced
the gas particles trapped in the center by stellar ones. 
This operation was repeated whenever the central gas accumulation was 
substantially slowing down the overall evolution.
We kept the softening of the individual particles when they were 
converted from one type to the other.
We have run a large number of tests to verify that this action did not affect
the evolution of the bar. This was achieved by running parallel models with
and without the gas particle replacement.


\subsection{Initial conditions and model parameters}
\label{s_inicon}

The initial conditions of the stellar and DM particles were created 
with the procedures described in Paper~I, using the density profiles 
from Hernquist (1993). The mass volume density distribution in the 
disk is given in cylindrical coordinates by

\begin{equation}
\label{eq:rho_disk}
\rho_\mathrm{d} (R,z) = \frac{M_\mathrm{d}}{4\pi h^2 z_0} 
      \exp (-R/h)\ \mathrm{sech}^2 \left(\frac{z}{z_0}\right),
\end{equation}
where $M_\mathrm{d}$ is the disk mass, $h$ is a radial scale length 
and $z_0$ is a vertical scaleheight. The density of the spherical halo 
is given by
\begin{equation}
\label{eq:rho_halo}
\rho_\mathrm{h}(r) = \frac{M_\mathrm{h}}{2\pi^{3/2}} \frac{\alpha}{r_\mathrm{c}} 
                        \frac{\exp(-r^2/r_c^2)}{r^2+\gamma^2},
\end{equation}
where $M_\mathrm{h}$ is the mass of the halo, $r_\mathrm{c}$ is a Gaussian
cutoff radius and $\gamma$ is the core radius. $\alpha$ is the normalization 
constant defined by
\begin{equation}
\alpha = \{ 1 - \sqrt{\pi} q \exp(q^2) [1-\mathrm{erf}(q)] \} ^{-1}
\end{equation}
with $q=\gamma/r_\mathrm{c}$. The particle velocities, dispersion velocities 
and asymmetric drift corrections were calculated using moments of the 
collisionless Boltzmann equation. Since models thus constructed are not in 
exact virial equilibrium, the halo component was relaxed for $t\sim 40$ in 
the frozen disk potential.

Because we are interested to quantify the effect of the gas fraction and
{\it gas} spatial resolution on the bar evolution, we use the pure stellar 
model SD from Paper~I as our benchmark model (Table~1). A fixed fraction 
\fg{} of stellar 
disk particles at $t=0$ were converted to identical mass gas particles 
and re-balanced using the central attraction forces from the total mass 
distribution. The gas is considered to be isothermal with $T_{\rm gas}=10^4$~K,
and initially moves on circular orbits.

We have created a set of models covering a two 
dimensional parameter space: by varying the gas mass fraction, \fg{}, in 
the disk, and by changing the limiting value of the gravitational softening, 
$\epsilon_{\rm dyn}$, in the gas. The sum of stellar$+$gas mass was kept constant in 
the models. We used the values 
$f_\mathrm{g} = 0\%$, 2\%, 4\%, 8\%, 15\%, 30\% and 50\%, and 
\egrav$ = 0.016$, 0.05 and 0.1. The softening used
for the stellar and DM components is fixed at 0.016.
Table~\ref{param} shows the combination of \fg{} and 
\egrav{} values used in each model.

\begin{table}[!h]
\centering
\caption{Parameters of the Standard Model}
\begin{tabular}{l r || l r}
\\
\hline
\multicolumn{2}{c||}{HALO} & \multicolumn{2}{c}{DISK} \\
Parameter & Value & Parameter & Value\\
\hline
$N_{\rm DM}$       &  $10^6$  &  $N_*$              &  $2\times 10^5$ \\
$M_\mathrm{h}$     &  3.15    &  $M_\mathrm{d}$     &  0.63 \\
$r_{\rm t}$        &  8.55    &  $R_{\rm t}$        &  1.71 \\
$\gamma$           &  0.1425  &  $h$                &  0.285 \\
$r_\mathrm{c}$     &  2.85    &  $z_\mathrm{0}$     &  0.057 \\
                   &          &  $Q_*$                &  1.5 \\
\hline
\end{tabular}
\tablecomments{$Q_*$ is the Toomre parameter for the stellar component fixed at 
$R=2.4h$, where $h$ is the thickness of the disk; $r_{\rm t}$ and $R_{\rm t}$ are 
numerical truncation radii in the halo and the disk. All values are given in 
dimensionless units, \S2.}
\label{sd_param}
\end{table}

\begin{table}[!h]
\centering
\caption{Gas fractions and limiting dynamical softening in the model sequences}
\begin{tabular}{l r r}
\\
\hline
Model    & \fg (\%) & ~\egrav \\
\hline
SD       &  0  &  --   \\
\\
SD\_G2S1   &    2 &  0.016  \\
SD\_G4S1   &    4 &  0.016  \\
SD\_G8S1   &    8 &  0.016  \\
SD\_G15S1  &   15 &  0.016  \\
SD\_G30S1  &   30 &  0.016  \\
SD\_G50S1  &   50 &  0.016  \\
\\
SD\_G2S2   &    2 &  0.050  \\
SD\_G4S2   &    4 &  0.050  \\
SD\_G8S2   &    8 &  0.050  \\
SD\_G15S2  &   15 &  0.050  \\
SD\_G30S2  &   30 &  0.050  \\
SD\_G50S2  &   50 &  0.050  \\
\\
SD\_G2S3   &    2 &  0.10  \\
SD\_G4S3   &    4 &  0.10  \\
SD\_G8S3   &    8 &  0.10  \\
SD\_G15S3  &   15 &  0.10  \\
SD\_G30S3  &   30 &  0.10  \\
SD\_G50S3  &   50 &  0.10  \\

\hline
\end{tabular}
\tablecomments{All values are given in dimensionless units, \S\ref{numerics}.
Columns (1) model sequences; (2) gas fractions in \%; (3) gravitational
softening in the gas.}
\label{param}
\end{table}

\section{Results}

Disks with a high content of gas tend to form dense clumps of gas which interact 
with the stellar component and raise the velocity dispersion in the disk. 
If this rise in the disk `temperature' is substantial, the bar instability 
may be lessened or completely suppressed (Shlosman \& Noguchi, 1993). 
We have encountered this problem when evolving models with \fg=100\% which
prevented us from completing these runs.

\subsection{Bar strength}

\begin{figure*}[!t]
   \centering
   \includegraphics[angle=270,scale=0.7]{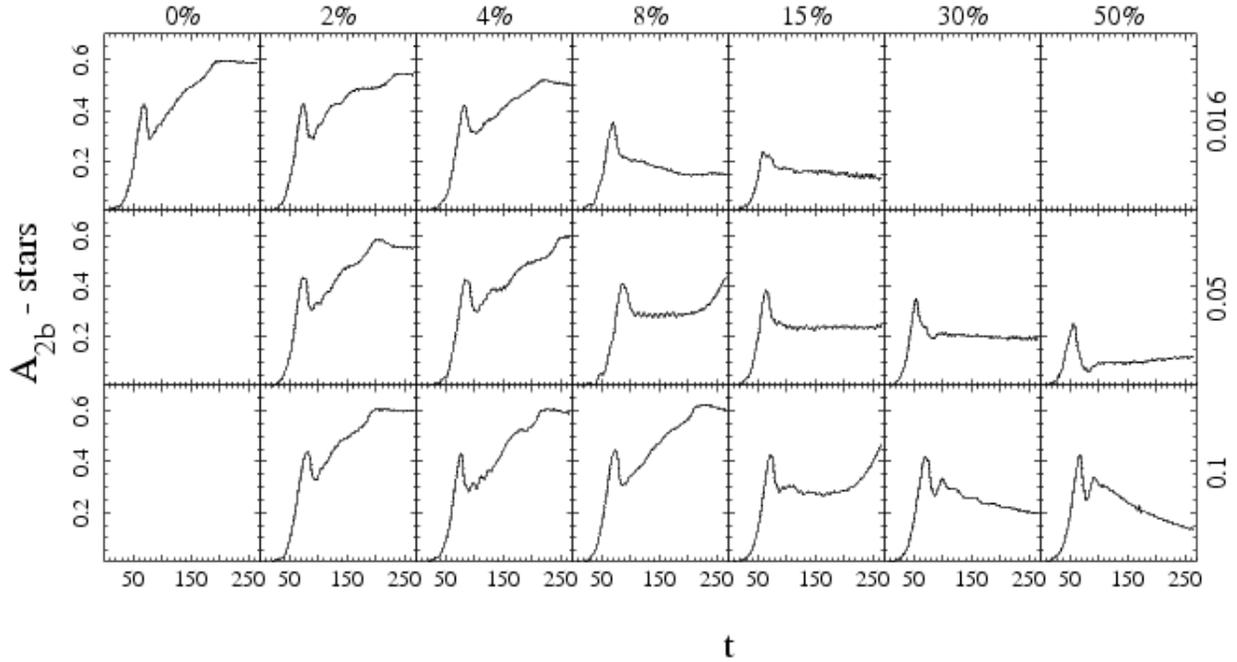}
   \caption{Time evolution of normalized bar amplitudes \Atworb{}.
     Rows correspond to models of equal dynamical softening in the gas, 
     \egrav{}, indicated on the right. Columns correspond to models with 
     equal gas fraction, \fg, indicated on the top.
     All the data has been smoothed with a lowpass Fourier filter.
             }
   \label{A2b}
\end{figure*}

The {\it stellar} bar strength has been quantified using the Fourier amplitude 
\Atworb{} of the $m = 2$ mode normalized by the $m = 0$ mode (Paper~I). 
It is obtained by integration over restricted cylindrical volumes where the
bar is a dominant morphological feature, namely, over $R = 0.1 - $\Rbar{} range,
where \Rbar{} is the bar size defined in Section 3.3 (see also Paper~I).
In Figure~\ref{A2b}, we plot \Atworb{} as a function
of time $t$ for various gas fractions, \fg, and gravitational softening in the
gas, \egrav.
The initial stages in the evolution of \Atworb{} are very similar in all models:
an initial stage of an accelerated growth, and a peak followed by a
sudden drop. The duration of this dynamical 
stage varies from model to model, but is completed by $t\sim 100$. 

The first peak in \Atworb{} is followed by the vertical buckling instability
in the bar leading to an abrupt weakening in the bar but not a complete
dissolution (e.g., Martinez-Valpuesta \& Shlosman 2004). We observe that
the bar weakening is quite independent of \fg{} and \egrav, as expected. 
For gas-poor models,
the first peak of \Atworb{} is independent of \egrav{} and \fg. For 
gas-rich models, with \fg$\gtorder 15\%$, the peak is lowered
gradually for \egrav $\ltorder 0.05$, up to a factor of 0.4. Models
with \egrav $=0.1$ appear not to be affected at all by this trend. 

The post-buckling evolution of the bars is much more affected by \fg{} and
\egrav, and shows a bimodal behavior. Namely, in gas-poor models,
the bar resumes its growth but at a more gradual pace as compared to the 
dynamical growth. In gas-rich models, the bar strength declines over
the simulation (i.e., Hubble) time. This bimodal behavior was noted by 
Berentzen et al. (2007), but the current models show it more explicitly. 

We shall refer to the first bar evolution phase, including the buckling,
as the {\it dynamical} phase, and to the subsequent evolution as the {\it secular} 
phase. The secular stage allows the separation of the secularly growing models  
from the secularly declining ones. In no models have the bars 
disappeared completely --- at least a substantial oval distortion remained.

The borderline \fg{} between these two trends, growth and decline, depends on 
\egrav. For \egrav$=0.016$, it lies in the range of \fg$\sim 5\%-7\%$. Larger 
\egrav{} moves the borderline \fg{} toward more gas-rich
models. Models SD\_G8S2 and SD\_G15S3 lie close 
to this borderline, \fg$\sim 8\%-10\%$ for \egrav$=0.05$ and \fg$\sim 10\%-12\%$ 
for \egrav$=0.1$, and show a very singular evolution --- 
after the sudden drop  \Atworb{}$\sim $const. for about $\Delta t\sim 100$, and
a second period of bar growth begins. This is a surprisingly mixed behavior 
showing the prolonged constancy in \Atworb{} of the gas-rich 
models and the secular bar growth of the gas-poor ones. 

Model SD\_G50S2 has a peculiar evolution in the dynamical stage that 
deserves special attention. Even though this is a gas-rich model,
the bar resumes is secular growth after the buckling. The rate of growth is 
much more moderate than that in the other models but is clearly noticeable.

\subsection{Pattern speed}

\begin{figure*}[!t]
   \centering
   \includegraphics[angle=270,scale=0.7]{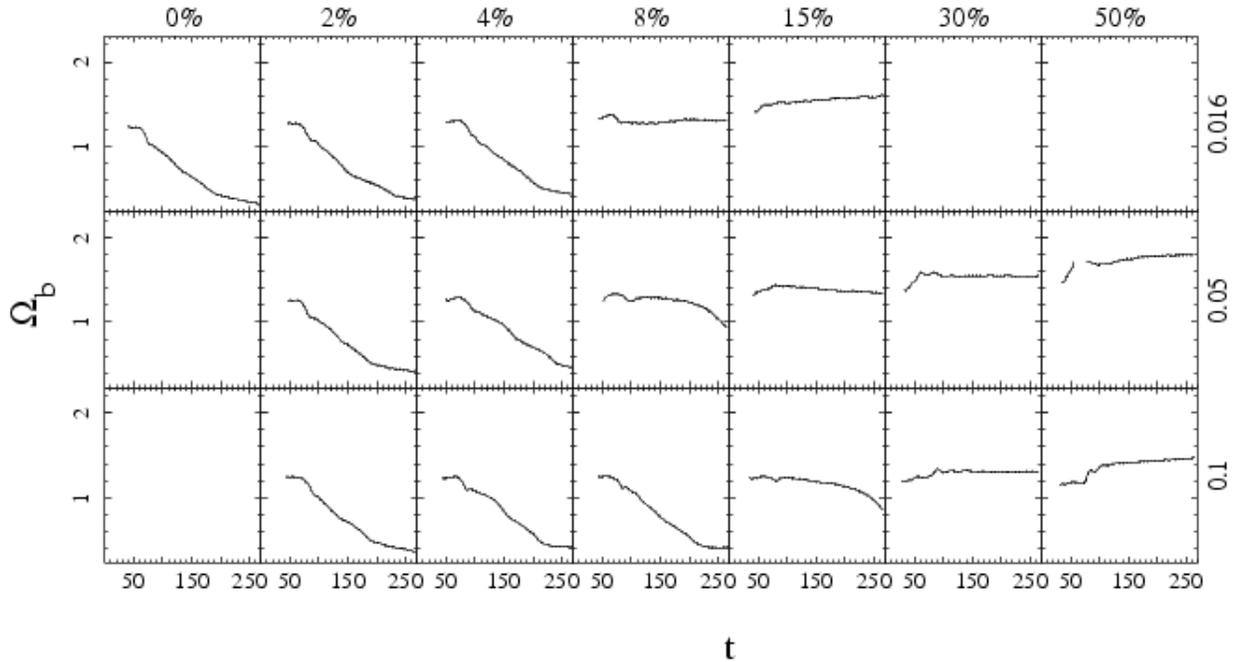}
   \caption{Time evolution of the bar pattern speed \ps{}.
     The distribution of models in rows and columns is that of 
     Fig.~\ref{A2b}. Model SD\_G50S2 curve has a gap around the time of the
     buckling. The bar weakens abruptly and its figure becomes distorted
     which makes it difficult to determine the pattern speed.
     Note that for the gas-rich disks, the pattern speed is a non-decreasing 
     function of $t$. The gas fractions and spatial resolution are
     indicated at the top and on the right respectively.}
     \label{figps}
\end{figure*}

The evolution of the pattern speed \ps{} is shown in Fig.~\ref{figps}. 
It remains about constant during the dynamical phase of the bar evolution,
slightly rising in the gas rich models. The bimodality of gas-poor and
rich models shows up in the secular evolution where \ps{} drops in
the former and stays constant or rises in the latter disks.
In the secular phase, \ps{} nearly always anticorrelates with \Atworb{} 
A sustained drop in \ps{} corresponds to a rise 
in \Atworb{}. Thus, in the gas-poor models, the bar tumbling slows down, 
while in the gas-rich models it stays nearly constant.

Fig.~2 provides a hint to what one can expect for the evolution of the
angular momentum in the disk-halo system, although in no way does this figure
account for {\it all} the angular momentum in the disk, and even in the
bar region it represents only the tumbling of the bar. The gas-poor 
disks lose their tumbling angular momentum
efficiently, while gas-rich ones nearly conserve it or even speed up their 
tumbling during the secular phase. 

\subsection{Bar length}

\begin{figure*}[!t]
   \centering
   \includegraphics[angle=270,scale=0.7]{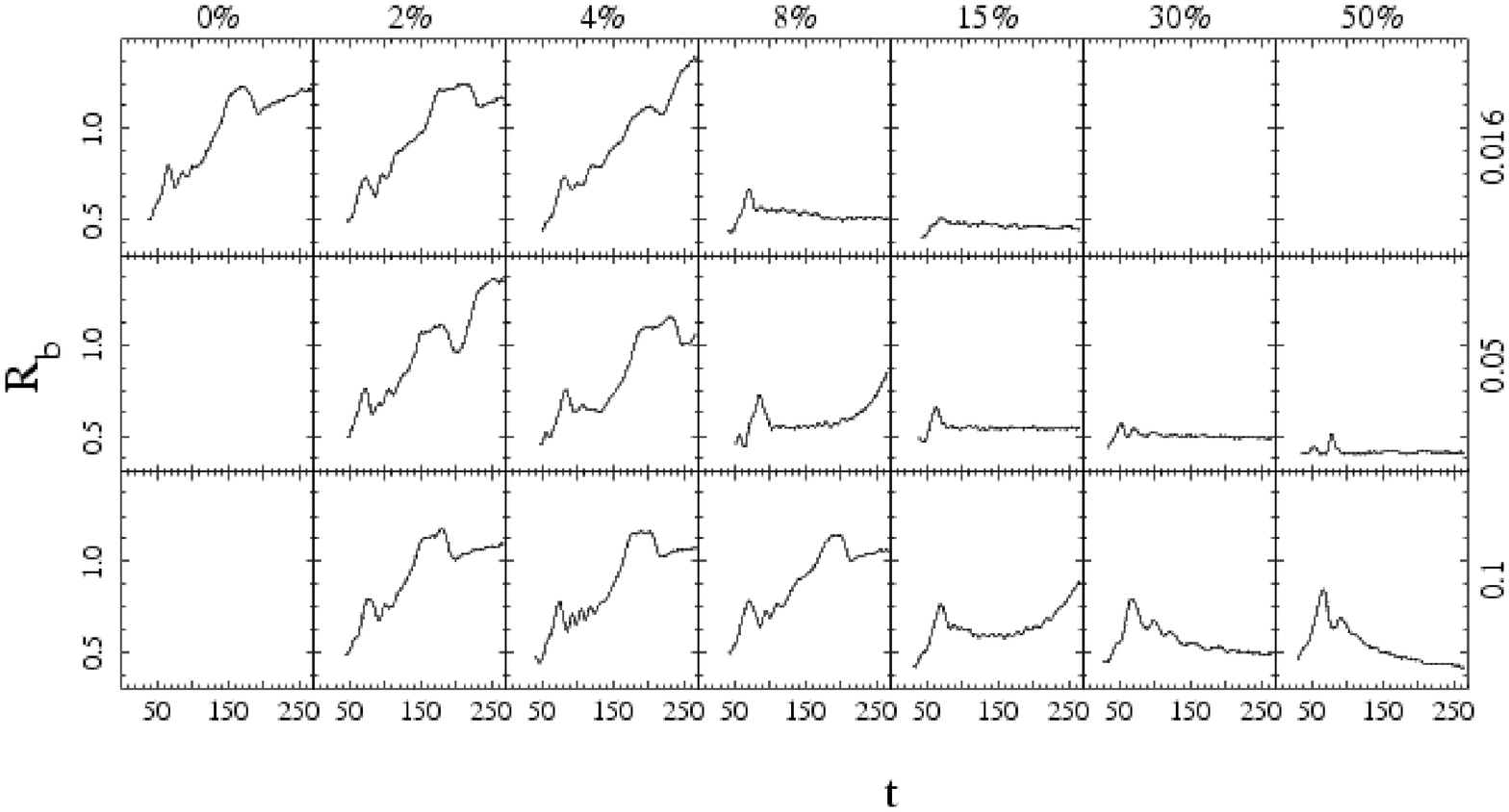}
   \caption{Time evolution of the bar semimajor-axis length \Rbar{}.
     The distribution of models in rows and columns is as in 
     Fig.~\ref{A2b}. The bar sizes include the {\it ansae} (see text).
     Note that bars are not growing in the secular phase
     in the gas rich disks (see also SN93), and this is more pronounced
     in high resolution simulations. Gas fractions and resolution are indicated 
     at the top and on the right respectively.}
     \label{figrbar}
\end{figure*}

The length of the bar semi-major axis, \Rbar, has been taken
as the radius where the bar equatorial ellipticity drops by 15\% off 
its peak (Paper~I). This method has been tested in comparison with 
alternative method based on the last stable orbit supporting the bar 
(Martinez-Valpuesta et al. 2006). It is the most reliable when applied 
after the first maximum of the bar strength. The ellipticity of the bar at
different radii is obtained by fitting ellipses to the isodensity
contours in the face-on disk. As in the collisionless models, the bar 
length exhibits the initial period of an accelerated growth and reaches 
a maximum which always coincides (in time) with the maximum in \Atworb{} 
(Fig.~\ref{figrbar}). This symbolizes the period of a dynamic
growth of the bar, or growth related to the bar instability itself.
Similarly to the pure stellar model, there is a subsequent drop in \Rbar{}
as a result of the vertical buckling (section 3.4). During the secular phase,
the bar length exhibits a more complicated behavior than 
in the collisionless models. Most importantly, it grows in gas poor
models while it stagnates or even shortens in the gas rich disks. 

\begin{figure*}[!t]
   \centering
   \includegraphics[angle=270,scale=0.7]{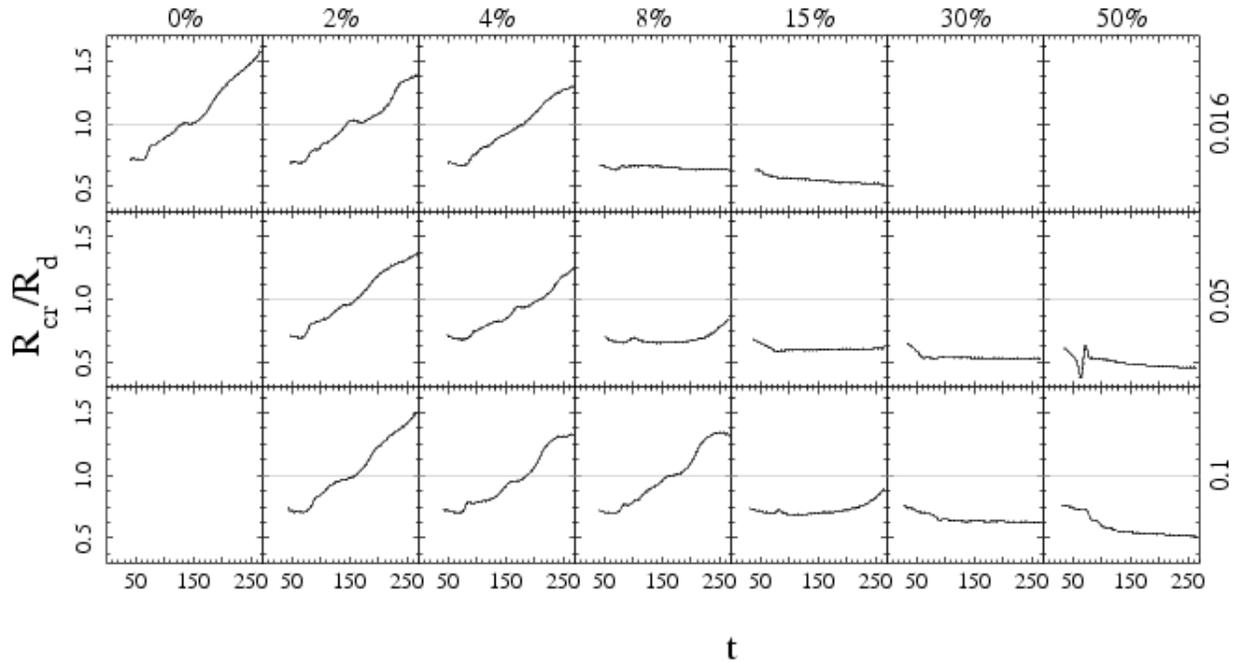}
   \caption{Time evolution of the ratio $R_\mathrm{cr}/R_\mathrm{d}$.
     The horizontal line has been added to help identify the 
     time \Rcr{} crosses the disk border.
     The distribution of models in rows and columns is as in 
     Fig.~\ref{A2b}. Note that for the gas-rich disks the CR is always inside
     the disk, especially for higher resolution disks. Gas fractions and 
     resolution are indicated at the top and on the right respectively.}
     \label{figRcrRd}
\end{figure*}

In the late stages of secular evolution, the gas poor models show a sudden
decrease in the bar size which has no immediate correspondence to \Atworb{}
evolution. Rather generally, 
the evolution of \Rbar{} does not go in tandem with changes in \Atworb{}. 
As an example, in model SD\_G4S1 after $t\sim 220$, \Rbar{} continues 
its growth while \Atworb{} growth saturates. This behavior is caused by
the formation, growth and detachment of the {\it ansae}\footnote{The ansae 
(i.e., handles) are typically found in early-type disk galaxies
(Fig.~11 in Martinez-Valpuesta et al. 2006; see also Martinez-Valpuesta, 
Knapen \& Buta 2007), e.g., NGC~4262, NGC~2859, and NGC~2950 
(Sandage 1961), NGC~4151 (Mundell \& Shone 1999), and ESO~509-98 (Buta et 
al. 1998).}, as discussed in Paper~I. In the following, we shall
show that \Rbar{} correlates with other properties of the bar-disk
system.

The bar corotation (CR) radius \Rcr{} has been computed using linear 
approximation. \Rcr{} grows or shrinks as a consequence of the variation 
of the bar tumbling speed, \ps. It grows substantially in the SD and gas 
poor models, while it stays 
constant or even drops slightly in the gas rich models. Its initial value 
among all models is the same. As in Paper~I, we follow the ratios of 
\Rcr{} to that of the disk \Rd{} (Fig.~\ref{figRcrRd}) and bar \Rbar{} sizes. 
Both \Rd{} and \Rbar{} have been
defined in Paper~I. Clearly, when \Rcr/\Rd$\sim 1$, its 
growth experiences a temporary slowdown and then resumes, 
rising to larger values. This effect is the result of the combination 
of a temporary slowdown of the outwards radial movement of \Rcr{} 
and a temporary expansion of the disk radial border which results 
from the drag on some of the disk material trapped at the CR resonance. 
As in the pure stellar models, the growth of the bar strength is sensitive
to the moment at which \Rcr{} crosses the edge of the disk, seen as 
a temporary period of slower growth of \Atworb{}. Note, that
\Rcr/\Rd{} is always less than unity in the gas rich
models and it seems that for a fixed \fg, the CR-to-disk
size ratio is damped stronger for small \egrav, i.e.,
for higher numerical resolution. 
 
The ratio \Rcr/\Rbar{} shows a pronounced dip during the
initial growth of the bar. This is followed by an equally abrupt 
rise due to the bar shortening that results from the vertical buckling 
instability. After these variations, both \Rbar{} and \Rcr{} grow 
in tandem, and \Rcr/\Rbar$\sim 1.5-2\pm 0.15$ in most of the models. 
Exceptions are
models SD\_G4S2 which has a rise and drop resulting from a temporary 
stall of \Rbar{}, and models SD\_G30S3 and SD\_G50S3 where \Rcr{}$\sim $
const. while the bar shrinks. Finally, a pronounced rise above 
\Rcr/\Rbar$=2$ occurs in models due to detachment 
of the ansae (Paper~I).
This rise is absent in models,
in which no ansae are formed or which form too late in the run and do not 
have time to detach from the bar, as in SD\_G15S3. We find that the ansae
do not form in models with more than $\sim 10\%$ of the gas. We note that
the bar size evolution given in Fig.~3, includes the ansae. However,
in Section~4, we find it advantageous to exclude the ansae when
discussing some correlations.

\subsection{Vertical buckling}

\begin{figure*}[!t]
   \centering
   \includegraphics[angle=270,scale=0.7]{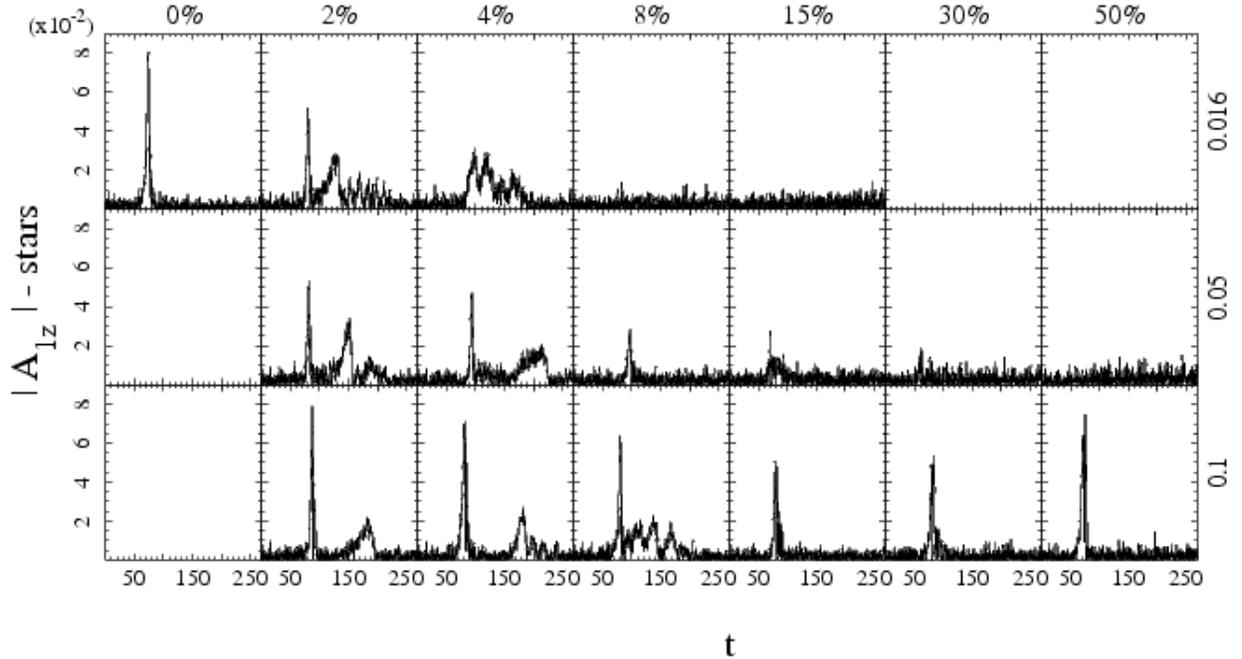}
   \caption{Evolution of the bar vertical asymmetry measured by the 
     Fourier coefficient $|A_{\rm 1,z}|$ of the $m=1$ mode in the 
     $rz$-plane corotating with the bar major axis. No filtering
     was applied to this data.
     The distribution of models in rows and columns is as in 
     Fig.~\ref{A2b}. Note that the first buckling amplitude decreases with \fg,
     but the sequence \egrav$=0.1$ exhibits a somewhat more complicated
     behavior. In addition, $|A_{\rm 1,z}|$ correlates with \egrav. 
     Gas fractions and 
     resolution are indicated at the top and on the right respectively.}
     \label{figA1z}
\end{figure*}

The vertical buckling observed in many numerical bars is an event 
that in some cases can reshape the phase-space density of the disk.
As we found in Paper~I, when the disk is heated vigorously by the 
buckling instability, the secular growth of the bar can be seriously 
diminished or completely halted. We have measured the vertical asymmetry 
of the 
stellar disk with the index \Aone{}, computed as described in Paper~I. 
The evolution of \Aone{} is shown in Fig.~\ref{figA1z}.
Most of the models have at least one clear peak between $t=50$ and 100, 
and this peak coincides in time with the drop in \Atworb{}. Exceptions are 
models SD\_G15S1, and SD\_G50S2, which have no clear peak 
above the noise level. Some models have secondary and even higher 
order peaks or prolonged periods where \Aone{} is clearly above the noise 
level, as found by Martinez-Valpuesta et al. (2006).

\begin{figure*}[!t]
   \centering
   \includegraphics[angle=270,scale=0.7]{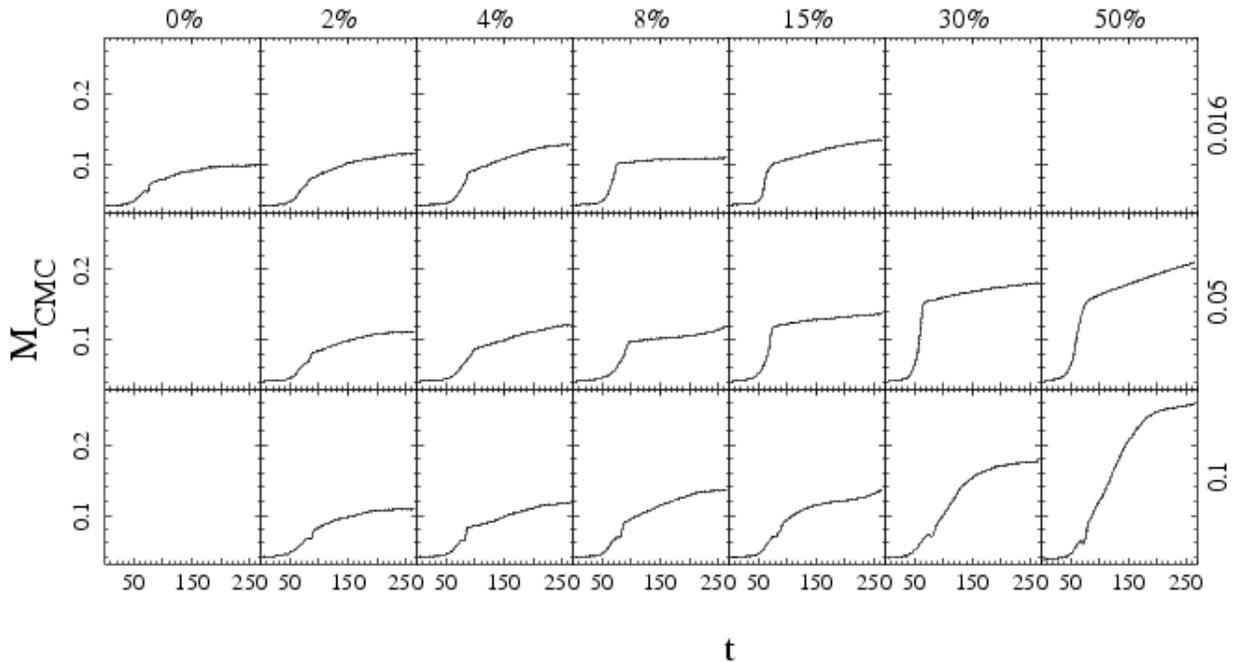}
   \caption{Time evolution of the central mass concentration, CMC: the total
     mass contained inside a sphere of 
     radius 0.1 at the center of mass of the stellar disk. The stellar, 
     gaseous and DM components are included in the mass count.
     All the data has been smoothed with a high frequency Fourier filter.
     Gas fractions and 
     resolution are indicated at the top and on the right respectively.
   }
   \label{figmtot1}
\end{figure*}

Clearly, the height of the first peak varies with the gas content of the disk.
In the sequences with \egrav$=0.016$ and 0.05, the peak becomes 
gradually lower in models with higher \fg. This result is in good 
agreement with Berentzen et al. (2007). The big surprise is in the 
sequence with \egrav$=0.1$, where the value of the peak 
drops as \fg{} increases from 0\% to 8\%, and then rises 
in models with \fg$=30\%$ and 50\%. This is discussed further in 
Section~\ref{s_disc}. 

The signature of the vertical buckling can be observed in the behavior of 
\ps. As the bar strengthens, it brakes against the outer disk and against
the DM halo, at later time and \ps{} declines. At the time of buckling,
the bar weakens abruptly and \ps{} experiences a break --- its slope
changes and becomes much smaller.

\subsection{Central mass concentration}

One of the differences that arise between pure stellar models of galactic 
bars and models including a gas component is the formation of a 
central mass concentration. 
The central densities, both stellar and DM, have notably increased 
even in pure stellar models as a response to the asymmetric potential of the bar
and the vertical buckling (Paper~I; Dubinski, Berentzen \& Shlosman 2009). 
However, this effect is by far more intense in models with gas.
The growth of central mass concentration (CMC) is boosted by the bar which 
channels the gas toward the central kpc.
This leads to a CMC composed mainly of gas, and, in
smaller proportion, of stellar and DM particles. In all our models, 
the CMC has resided inside a radius smaller than 0.1. 

To follow the growth of the CMC, we have measured the mass \mcmc{} contained 
within a sphere of radius 0.1 placed at the center of mass of the disk. 
Contributions from all three components have been included (Fig.~6). 

\begin{figure*}[!t]
   \centering
   \includegraphics[angle=270,scale=0.7]{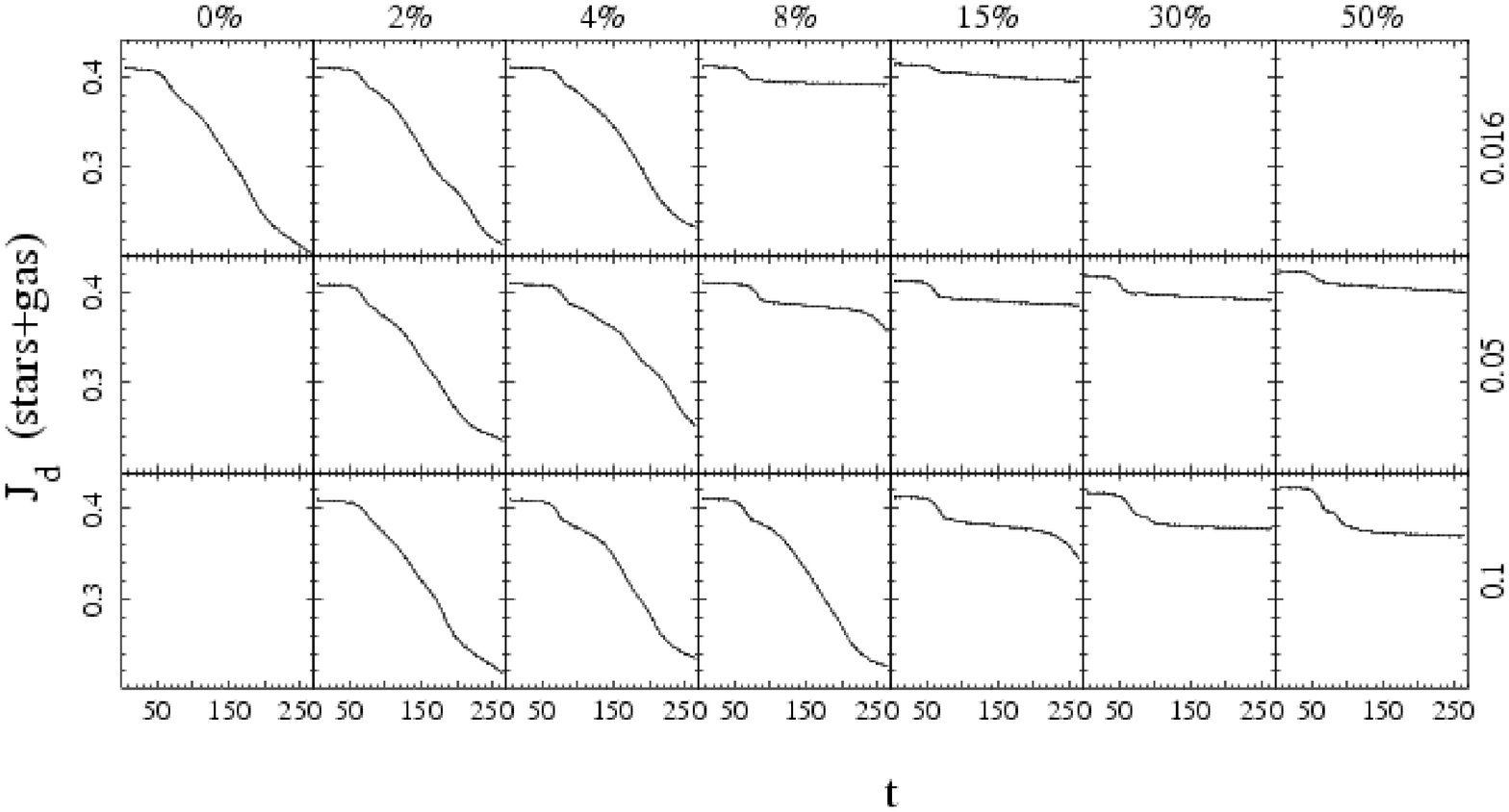}
   \caption{Time evolution of disk total angular momentum, $J_{\rm d}$.
      Gas fractions and 
     resolution are indicated at the top and on the right respectively.
   }
   \label{figmtot1}
\end{figure*}

Model SD\_G2S1 serves as a benchmark of evolution in the gas poor models. 
With the formation of the stellar bar, a massive accumulation of 
gas appears  at the center, with an elongated shape of $\sim 0.1\times 0.04$,
in the disk plane. This CMC is approximately aligned with the 
major axis of the (gas) bar. It rapidly captures the gas fed by the bar 
and simultaneously contracts into a small, more axisymmetric and somewhat 
flattened circular `blob' with $\Delta R\sim 0.04$ and $\Delta z\sim 0.01$.
These characteristic sizes are determined largely by \egrav.
The accelerated \mcmc{} rise seen in Fig.~6 encompasses the 
exponential bar growth phase, up to the time of buckling (or the time when 
\Atworb{} drops in models with no buckling). 
Typically, about 50\% of the gas mass is captured by the CMC during this 
stage. 

During the secular phase of the bar evolution, the CMC captures the gas
at a much slower pace. This is mostly the gas which remains outside \Rbar{}
and even \Rcr{} at the onset of the secular phase.
The slowing down of the CMC growth results from a combination 
of two factors. First, the availability of the disk gas has been  
severely reduced by the violent initial inflow, and second, 
the bar strength has been diminished by the buckling.
As a consequence, the gas-poor models have been left with very little gas in the 
disk after the initial inflow, although they show a healthy secular bar growth.
Whereas gas-rich models have gas left, but their bars are weak and do not
grow, so no fresh gas crosses \Rcr{} which stagnates as well. 

Three models exhibit a somewhat different evolution for \mcmc.
Instead of a short period of a very accelerated growth, the CMC grows at 
a somewhat slower rate but over an extended period of time. 
These are models SD\_G15S3, SD\_G30S3 and SD\_G50S3, which are all gas-rich and
have the lowest resolution. 
In these models, the gas content of the bar has formed without a visibly 
prominent CMC. 
Instead of a rapid influx to the center, the gas in the bar 
contracts rather slowly, gradually increasing its density at the center.
The process saturates when the gas is completely contained within the 
central $R\sim 0.1$.

Models SD\_G8S2 and SD\_G15S3 show a rate of \mcmc{} growth that increases 
with time at the end of the run. This behavior  
is related to the late period of accelerated bar growth.

\subsection{Angular Momentum}

\begin{figure*}[!t]
   \centering
   \includegraphics[angle=270,scale=0.7]{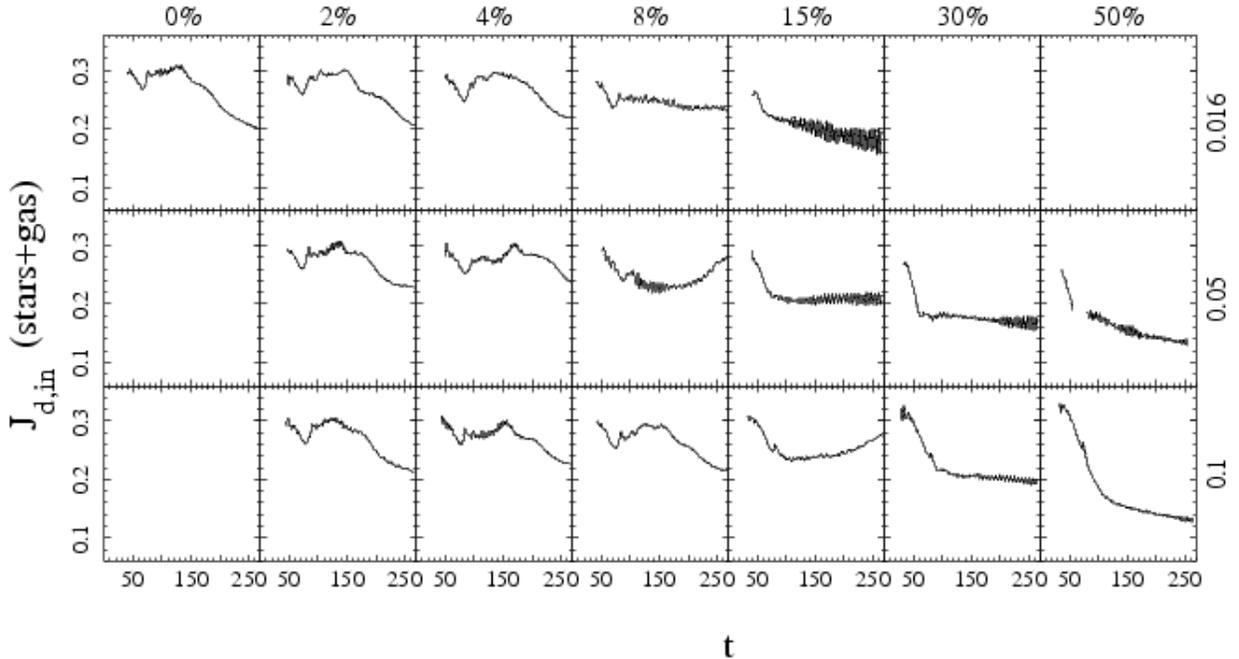}
   \caption{Time evolution of disk angular momentum inside 
            the bar corotation, \Jdin. Note that \Jdin$\approx $const. in
            time as long as \Rcr{} remains inside the disk, as can be seen
            from Fig.~4. Gas fractions and 
     resolution are indicated at the top and on the right respectively.
   }
   \label{figmtot1}
\end{figure*}

Angular momentum evolution in the disk has been followed within a number of 
characteristic radii, namely, the CR (\Jdin) and the disk radius, \Jd{}
(which contains 98\% of the disk mass by definition). The latter exhibits 
a clear evolutionary sequence as a function of \fg{} (e.g., Fig.~7). Both
the amount of $J$ lost by the disk over dynamical and secular evolution
decreases monotonically from pure stellar disks to progressively more 
gas-rich ones. In a way this is a reflection of weaker bars along \fg.
Even more interesting is the evolution of \Jdin{} (Fig.~8). After the 
initial adjustment,
the disk within the CR is losing its angular momentum. The loss is not
dramatic, $\sim 10\%-20\%$, and somewhat increases with \fg. This decline in \Jdin{}
continues until the vertical buckling sets in. For low \fg{} disks, the buckling 
is associated
with a nearly sudden increase in \Jdin{} which restores the pre-buckling
value followed by a subsequent gradual
decline. Gas-rich disk show no sudden increase in \Jdin{}
but rather a slow decline.
 
\Jdin{} stays quite flat until $t\sim 150$, a long time after the buckling,
as can bee seen from Fig.~8 (with the exception of a small initial decline and
increase, as mentioned above). Interestingly, this time corresponds to 
\Rcr/\Rd$ < 1$ in the gas-poor models (Fig.~4). Crossing \Rcr/\Rd$ = 1$
border
affects the bar growth (Fig.~3), which saturates immediately. It also is reflected
in the evolution of \Rcr/\Rbar{} which increases abruptly thereafter. Clearly,
the bar growth ceases at some point when the \Rcr{} exceeds \Rd{} and the bar
cannot capture additional orbits and be fed by the angular momentum from these
orbits. This situation is similar to that analyzed in Paper~I (Section~3.6),
where the angular momentum of the disk inside \Rcr{} stays about constant 
as long as \Rcr{} remains within the disk. Our present models can be directly
compared to the Standard model of Paper~I. We return to this point in Section~4. 

The loss of the angular momentum, \Jdout, with \fg{} by the outer disk appears to 
be much more severe than \Jdin. But one should remember that the outer disk,
i.e., outside the CR, accounts for less mass. Before the buckling, \Jdout{}
behaves similarly to \Jdin, but differently at the later times. The trend 
here is that \Jdout{}
declines steeply with time for gas poor disks after the buckling period.
It stays nearly constant with time for the gas rich models.
 
Models with different \egrav{} show a behavior consistent with our 
understanding. There is a slight increase in the $J$ transfer during dynamical 
and secular stages of the bar evolution. Much less difference is observed in 
the evolution of \Jdin{} and \Jdout{} in this case.

\section{Discussion: Testing New Correlations}
\label{s_disc}

\begin{figure*}[!t]
   \centering
  \includegraphics[angle=0,scale=1.0]{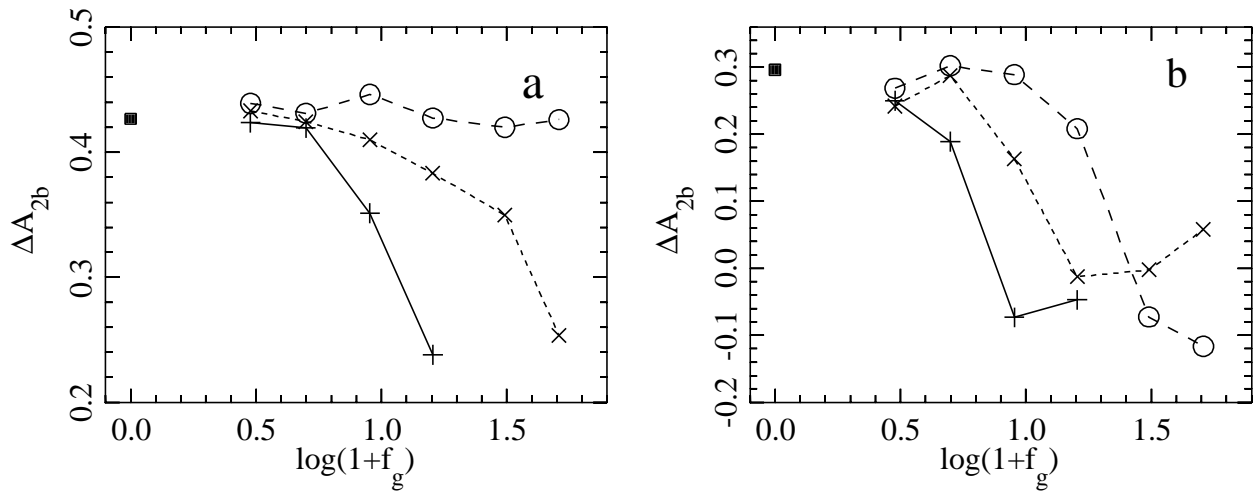}
   \caption{Bar strengthening during $(a)$ dynamical and $(b)$ secular phases of 
   its evolution as a function of the gas fraction, \fg, for different spatial
   resolution in the gas, \egrav, namely \egrav$=0.016$ (crosses, solid lane),
   0.05 ($\times $, short dashed lane), and 0.1 (circles, long dashed lane).
   The filled square represents the SD model (Paper~I) with no gas. The
   model SD\_G8S1 has been omitted (see the text).
   }
   \label{figmtot1}
\end{figure*}

We have studied the effects of gas fraction, \fg, and spatial resolution, \egrav, 
on some aspects of bar evolution in galactic disks embedded in DM halos. 
Specifically, we aimed at understanding their effect on the basic parameters 
characterizing stellar bars, e.g., bar size, corotation radius, strength, etc. 
We have also followed the angular momentum redistribution in the disk-halo 
systems as a function of \fg{} and \egrav. The SD pure stellar model
(Paper~I) has been used as a template --- the disk and halo parameters in current
models have been fixed at those of the SD. As before, the bar evolution has been 
separated into two main phases --- the dynamical phase, which represents the bar 
instability itself de facto, and the secular
phase. These phases are separated by the vertical buckling instability in the bar. 
We investigate a number of new correlations which are corollaries
to evolution discussed in section~3.
In the follow-up work (Villa-Vargas et al., in preparation), we analyze the 
angular momentum transfer and bar evolution 
as a function of disk gas fraction and by varying the basic parameters of the
DM halo. 

Overall, there appears to be a substantial difference in the evolution of the 
gas-poor and rich models, as has been shown in the previous section. This is 
discussed below in more detail. The boundary
between the gas-poor and gas-rich models depends on the spatial resolution 
for the gas component --- it lies around 5\%--7\% for \egrav=0.016, and shifts
to $\sim 10\%-12\%$ for \egrav=0.1. The loss of spatial resolution, in a way, 
changes the nature of the gas component --- it becomes less dissipative.  We 
define the bar strengthening during
its dynamical phase as $\Delta $\Atworb~$\equiv $\Atworb ($t=t_{\rm peak}$)\
--\Atworb($t=0$), where $t_{\rm peak}$ is the time when the bar has reached 
its maximal amplitude prior
to the onset of buckling; the second term \Atworb($t=0$)$=0$. Similarly, 
the bar growth during the secular
phase is defined as $\Delta $\Atworb~$\equiv $\Atworb($t=270$) -- 
\Atworb ($t=t_{\rm min}$), where $t_{\rm min}$ is the time of the minimal
bar amplitude immediately following the buckling.

In the {\it dynamical} stage of the bar evolution, the largest differences 
have been observed in the gas-rich models, where $\Delta $\Atworb{} 
drops dramatically with \fg{} for \egrav$=0.016$ and 0.05, while it exhibits 
no dependence on \fg{} for 
\egrav$=0.1$ (Figs.~1, 9a). In the gas-poor models, the maximal bar amplitude  
achieved in the dynamical phase is nearly independent of the gas fraction, nor 
does it depends on the spatial resolution used in the modeling of the gas component 
(Figs.~1, 9a). Fig.~9a supports the Berentzen et al. (1998, 2007) analysis
that the increasing presence of the gas component makes the bar instability
milder. In the {\it secular} stage, this sharp decrease in $\Delta $\Atworb{}
with \fg{} persists, including 
the lowest resolution sequence as well. The bar growth is severely restrained 
in the gas-rich models in this phase. It either had vanished or even becomes
negative. The underlying physical process has been quantified by Berentzen et al. 
(1998): the {\it gas affects the orbital structure} of the collisionless components, 
which in turn modifies the bar properties. However detailed understanding of
how this influences particular parameters of the bars requires more work.

We have attempted to clarify this dependence of the bar strength on 
\fg{} and \egrav. As a working hypothesis, we have tested whether 
the final mass of the CMC, \mcmc, in each model can serve as an underlying 
hidden parameter replacing \fg{} and \egrav. The dependency of the final
bar strength on the CMC mass in the pure stellar disks has been analyzed
by Athanassoula, Lambert \& Dehnen (2005). Extensive study of the bar evolution
in the presence of an analytical CMC has been performed by Shen \& Sellwood (2004).
Bornaud et al. (2005) have attributed the bar weakening to both the CMC and the
angular momentum transfer from the gas to the stellar bar, using analytical
CMC and DM halo. However, the latter 
process has been found not important by Berentzen et al. (2007) following a
detailed analysis of the angular momentum transfer in live potentials of the
CMC and DM halo. Moreover, Berentzen et al. (1998, 2007) have 
used self-consistently growing CMCs in the gaseous/stellar disks to
demonstrate that the secular bar growth is strongly affected. 
The set of numerical models analyzed here is in fact the most controlled
experiment performed so far to test the influence of gas fraction and
its resolution on the bar evolution. 

\begin{figure*}[!t]
   \centering
  \includegraphics[angle=0,scale=1.0]{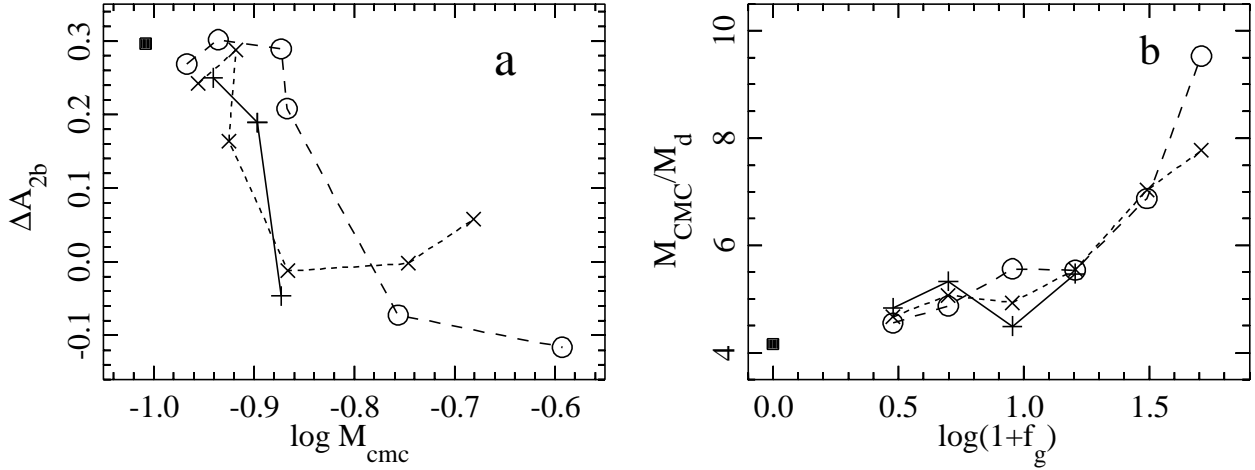}
   \caption{$(a).$ The secular strengthening of the bar, $\Delta$\Atworb, after 
   the first buckling and until the end of the simulations,
   as a function of the final central mass concentration CMC within $R=0.1$, 
   \mcmc{}, for 
   different spatial resolution in the gas, abbreviated as in Fig.~9. The 
   filled square represents the SD model (Paper~I) with no gas. The
   model SD\_G8S1 has been omitted (see the text).
   $(b).$ Final \mcmc{} normalized by the total mass of the disk at $t=0$ 
   as a function of \fg.
   }
   \label{figmtot1}
\end{figure*}

\begin{figure*}[!t]
   \centering
   \includegraphics[angle=0,scale=1.0]{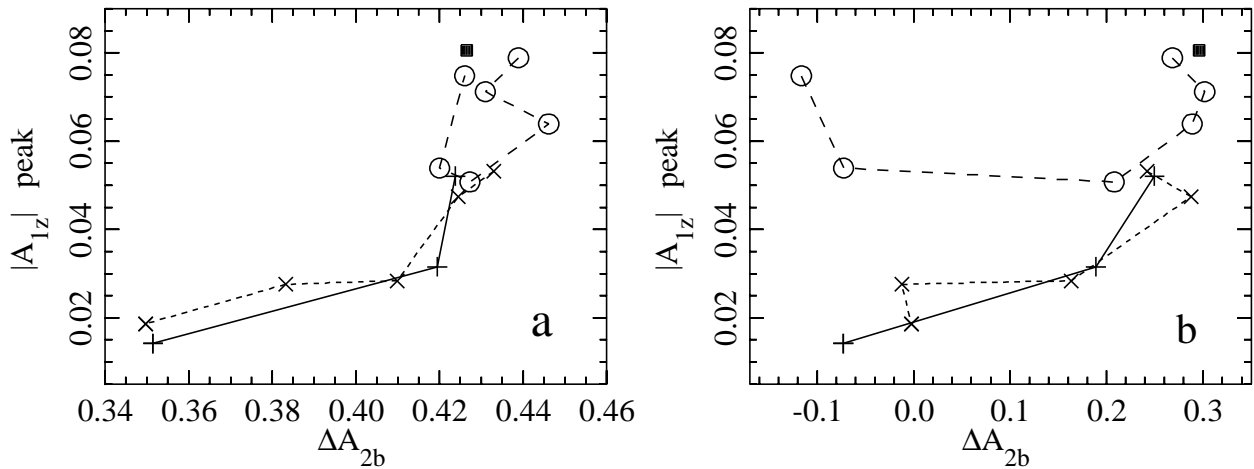}
   \caption{First vertical buckling amplitude of the stellar bar, $A_{\rm 1,z}$, as 
   a function of the bar strengthening during dynamical and secular phases of its 
   evolution, $\Delta $\Atworb, for different spatial
   resolution in the gas, abbreviated as in Fig.~9. The filled square represents 
   the SD model with no gas.
   }
   \label{figmtot1}
\end{figure*}

Fig.~10a shows the dependence of the bar strengthening after the first buckling, 
$\Delta$\Atworb{} on final \mcmc{}, with all 
other parameters characterizing the stellar disk and the DM halo being 
fixed. One model, SD\_G8S1, has been omitted,
as its CMC secular evolution is unusually flat and falls out of the
sequence with \egrav$=0.016$. If \mcmc{} is the single defining
variable which controls $\Delta$\Atworb, all three \egrav-sequences are expected to
merge into a single sequence in Fig.~10a. What is actually observed is 
that the three sequences
exhibit a very similar behavior --- each curve is flat for the gas-poor 
models, experiences an abrupt drop, and flattens out.
However, the curves do not coincide completely as would be expected if
\mcmc{} were the {\it unique} underlying parameter. In fact, in some
cases the points corresponding to the same \mcmc{} but different \egrav{}
have a substantial
vertical dispersion in $\Delta$\Atworb{}. Hence, the issue
remains inconclusive. In contrast, we show the final \mcmc{} value
normalized by the disk mass at $t=0$ as a function of \fg{} (Fig.~10b).
This fractional \mcmc{} value is indeed unique for the three sequences, as
all three curves have nearly merged.

We now turn to a plausible correlation between the buckling amplitude
and the bar strengthening. The vertical buckling (e.g., Toomre 1966; 
Combes et al. 1990; Raha et al. 1991) is a recurrent instability 
(Martinez-Valpuesta et al. 2006) that was recently 
analyzed by Berentzen et al. (2007) in the presence of gas (see also 
Berentzen et al. 1998). The gas component, it has been concluded, leads to
a milder instability. Here we have attempted to relate the buckling amplitude,
$A_{\rm 1,z}$, to the change in the bar amplitude, $\Delta $\Atworb, in the
dynamical and secular phases of evolution (Fig.~11). In both phases
we observe a clear correlation between \Aone{} and
$\Delta $\Atworb{}. Stronger bar instability leads to a stronger buckling,
and indeed, increasing \fg{} makes the bar and buckling instabilities 
milder (Figs.~11a;
see also Fig.~9a). On the other hand, stronger buckling goes in tandem
with the bar secular growth (Fig.~11b). This trend shows saturation for 
the strongest
bars. Higher resolution sequences with \egrav$= 0.016$ and 0.5 behave in a very
similar fashion, while the lowest resolution sequence stands out of this
correlation. We have also checked the value of the drop in \Atworb{}
during the buckling and find a clear match between the higher resolution
sequences, while the lower resolution models behave differently. 

The physical extent of the bar depends on its ability to capture
additional orbits. While in principle this capture can proceed at all
radii, the fertile region lies between the bar's end and its CR radius,
where various families of orbits can be easily destabilized.
Therefere, the  bar growth due to the orbit capture should go in tandem
with the angular momentum {\it influx} because the near-CR orbits will
have a larger momentum-to-energy, J/E, ratio than the bar orbits.
In Paper~I, we have shown, and this is confirmed here, that the influx of 
angular momentum across
the CR and into the bar is able to maintain \Jdin$\sim $const. in time,
as long the CR radius lies within the disk, but in the presence of the gas
{\it this statement is limited to the gas-poor models only}. This happens 
despite that the CR remains within the disk at all times and for all 
gas-rich models (Fig.~4). In other words, in gas-rich models the $J$ influx
across the CR cannot compensate for its loss within the CR. 

The amplitude \Atworb{} is expected to be related to the ability
of the bar to capture additional orbits in the bar CR region, leading
to its geometrical growth. In both the dynamical and secular phases of the
bar evolution we find that the final bar amplitude, \Atworb, correlates
with the final bar size\footnote{The only exception appears to be the low
resolution sequence in the dynamical stage whose points cluster in the
same region (Fig.~12a)} (Fig.~12). So stronger bars appear to be longer as well. 
Taken at face value, this property of bar evolution appears to be supported
by recent observations in that the $K$-band images of barred galaxies exhibit
a correlation between the bar amplitude, measured by the $m=2$ Fourier component,
and its size (Elmegreen et al. 2007). However, there are caveats.

\begin{figure*}[!t]
   \centering
   \includegraphics[angle=0,scale=0.9]{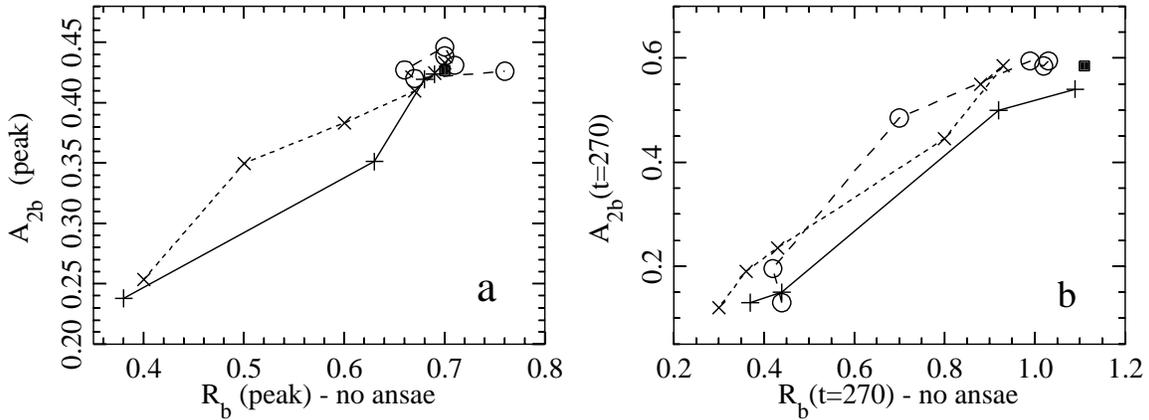}
   \caption{The bar amplitude, \Atworb, at its pre-buckling peak (left) and
   at the end of the simulation (right) as
   a function of the bar size (radius), taken at the same time, during
   dynamical and secular phases of the bar evolution.  Different spatial
   resolution in the gas has been abbreviated as in Fig.~9. The full square
   represents the SD model with no gas. We have subtracted the {\it ansae}
   from the bar sizes given in Fig.~3 (more details in the text).
   }
   \label{figmtot1}
\end{figure*}

\begin{figure}[!t]
   \centering
   \includegraphics[angle=0,scale=1.0]{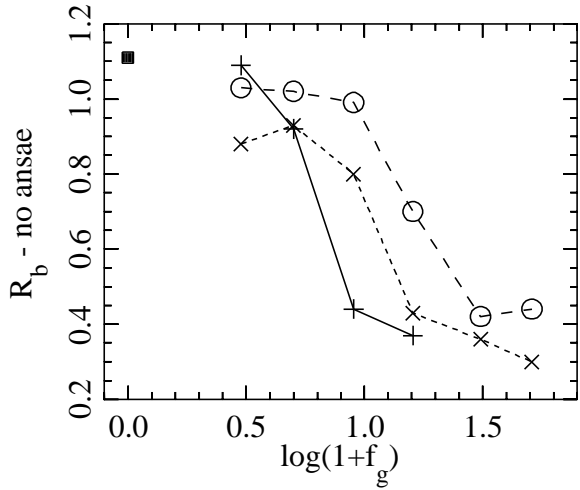}
   \caption{Bimodal distribution of the final bar sizes, \Rbar($t=270$), as
   a function of the gas fraction, \fg.  We have subtracted the {\it ansae}
   from the bar sizes given in Fig.~3 (more details in the text).
   }
   \label{figmtot1}
\end{figure}

We find that the bar sizes are substantially smaller in the gas-rich models
compared to the gas-poor ones (Figs.~3, 13). While the median bar size
at the end of the simulation in the latter models appears to be $\sim 1.2$, it 
is around 0.5 in the former ones, i.e., 12~kpc vs 5~kpc, embedded in the 
identical disks and halos. This difference arises almost entirely during the 
secular phase of the bar evolution. More precisely, by the end of the dynamical
stage, one can notice that the bar size slightly anti-correlates with \fg{} for
the gas-rich models and only for \egrav$=0.016$ and 0.05 sequences. This trend is
sufficiently weak and bars differ by not more than 10\%--15\% of their length.
So essentially, the bar growth in the dynamical phase is independent of \fg.

What is rather striking, is the abrupt change in the bar's growth `habit' in 
the secular phase --- bars grow in the gas-poor models and stagnate in the 
gas-rich ones. This leads to a {\it bimodal} evolution of the final bar sizes with 
respect to the gas mass fraction (Fig.~13). We note that this 
diverging evolution should be 
observable only in the long-lived bars which naturally reside in the older disks.
The relevant evolutionary timescale is that of a few Gyrs. The Milky Way disk is
a good candidate, as it is probably about 10~Gyr old and was not affected 
by major mergers over this time period (e.g., Gilmore, Wyse \& Norris 2002).

The anti-correlation between the final bar sizes and disk gas fractions, found 
here, complicates the simplistic picture of bar evolution.
The reason for this is that our gas-rich models lead to more massive CMCs
which increase bulge-to-disk mass ratios and, therefore, should be associated with 
earlier type disks. This means that smaller bars in our models lie in 
early-type disks (Fig.~13), while they are expected to be found in late-type 
disks, as noted by a number of surveys (e.g., Erwin 2005; see also Laine et al. 
2002). The
simplest resolution of this discrepancy can lie in the omission of star
formation processes and especially the feedback from stellar evolution and from
the central
supermassive black holes in this work. This leads to a gross over-estimate of the
amount of gas which reaches the central kpc and hence contributes to the
growth of the CMC by dragging stars and even DM inwards. We note, however, that the
purpose of this numerical exercise was to understand the effect of gas fraction
and gas spatial resolution on the disk-bar-halo interactions in the system.
Therefore, we have simplified the long list of processes known to affect
the galaxy evolution at large.

In summary, we find that the spatial resolution in the gas component becomes 
increasingly important for the bar evolution in the gas-rich disks. This is true 
for the dynamical but especially for the secular phase of evolution. In most
cases, model sequences with \egrav$= 0.016$ and 0.05 show a similar behavior, 
while differing substantially from the sequence with \egrav$=0.1$.

A bimodal behavior has been found for models based on their gas fractions.
The border line between the gas-poor and gas-rich systems appears to lie
around $5\%-7\%$ for higher resolution models. It shifts to $\sim 10\%-12\%$
for the lowest resolution models. The switch from a gas-poor to a gas-rich 
behavior appears to be sufficiently abrupt. It is clearly visible in all basic
characteristics of bar evolution, such as the bar strength, the CMC mass, the 
bar buckling amplitude, the bar size, etc. The largest differences in the 
evolution have been found in the secular phase.

We find that the presence of the gas component severely limits the bar growth
and affects its pattern speed evolution. While pure stellar models (Paper~I)
exhibit a rapid slowdown of the bar tumbling, as known for a long time 
(e.g., Debattista \& Sellwood 1998; Athanassoula 2003), the addition of a
substantial amount of gas reverses this trend completely.
Furthermore, the CR-to-disk size ratio, \Rcr/\Rd, was determined to be an important
dynamic discriminator between various phases of barred disk evolution.
Here we find that the gas-rich models are characterized by \Rcr/\Rd~$ < 1$ and
by $\Omega_{\rm b}\sim$~const. In these models, $\Omega_{\rm b}$ can even 
slightly increase with time. In 
addition, the gas-rich models maintain \Rcr/\Rbar~$ < 2$ which is much more
in agreement with observations of stellar bars being fast rotators.

The angular momentum evolution displays the same degree of bimodality with the 
gas fraction. For low \fg, the total disk $J$ drops steeply and monotonically
with time after the buckling, while it decreases weakly for the gas-rich models.
The reason for this behavior is of course that the bar amplitude is substantially
lower in the gas-rich models. Our attempt to explain this difference between the
gas-poor and rich models in terms of the more massive CMCs in the latter ones
has been rather inconclusive. We shall return to this issue in the forthcoming
work. 

Next, we have confirmed our previous claim (Paper~I) that the angular momentum,
\Jdin, within the CR radius is maintained at a constant level due to the
influx of angular momentum across the CR as it expands. We have extended this 
statement to the gas-poor models. For the gas-rich models, \Jdin{} drops abruptly
to a lower level and stays constant thereafter. 

A number of corollaries follow from the above results. We find that the bar 
strength inversely correlates with the gas fraction, both in the dynamical
and secular phases of bar evolution. The only exception seems to be the lowest 
resolution sequence in the dynamical phase which fails to capture this trend.
We also find that the buckling amplitude becomes larger for stronger bars
prior to the onset of buckling. On the other hand, the secular growth is most 
prominent in bars which show a large buckling amplitude. Finally, we show that
stronger bars are also the longest ones throughout both evolutionary phases,
and that bar sizes anti-correlate with the gas fraction.

\acknowledgments
We are grateful to our colleagues for numerous discussions. This research has 
been partially supported by NASA/LTSA/ATP/KSGC and  the NSF grants to I.S.
C.H. acknowledges the NSF grant.



\end{document}